\documentclass[aps,prb,reprint,superscriptaddress,showpacs, showkeys, showpacs]{revtex4-2}
\usepackage{hyperref}
\usepackage{graphicx}
\usepackage{amsmath}
\usepackage{amssymb}
\usepackage{color}
\usepackage{upgreek}
\usepackage{bm}
\usepackage{siunitx}
\begin{document}

\title{Strong coupling between WS$_2$ monolayer excitons and a hybrid plasmon polariton at room temperature}
\author{Yuhao Zhang}
\email{yuhaozhang@physik.uni-bonn.de}
\affiliation{Physikalisches Institut, Rheinische Friedrich-Wilhelms-Universität Bonn, 53115 Bonn, Germany}

\author{Hans-Joachim Schill}
\affiliation{Physikalisches Institut, Rheinische Friedrich-Wilhelms-Universität Bonn, 53115 Bonn, Germany}
\affiliation{Electron Microscopy and Analytics, Center of Advanced European Studies
	and Research (caesar), 53175 Bonn, Germany}
\author{Stephan Irsen}
\affiliation{Electron Microscopy and Analytics, Center of Advanced European Studies
	and Research (caesar), 53175 Bonn, Germany}
\author{Stefan Linden}
\email{linden@physik.uni-bonn.de}
\affiliation{Physikalisches Institut, Rheinische Friedrich-Wilhelms-Universität Bonn, 53115 Bonn, Germany}

\date{\today}

\begin{abstract}
Light-matter interactions in solid-state systems have attracted considerable interest in recent years. 
Here, we report on a room-temperature study on the interaction of tungsten disulfide (WS$_2$) monolayer excitons with a hybrid plasmon polariton (HPP) mode supported by nanogroove grating structures milled into single-crystalline silver flakes.
By engineering the depth of the nanogroove grating, we can modify the HPP mode at the A-exciton energy from propagating surface plasmon polariton-like (SPP-like) to localized surface plasmon resonance-like (LSPR-like).  
Using reflection spectroscopy, we demonstrate strong coupling between the A-exciton mode and the lower branch of the HPP for a SPP-like configuration with a Rabi splitting of 68 meV. 
In contrast, only weak coupling between the constituents is observed for LSPR-like configurations.
These findings demonstrate the importance to consider both the plasmonic near-field enhancement and the plasmonic damping during the design of the composite structure since a possible benefit from increasing the coupling strength can be easily foiled by larger damping.


\end{abstract}

\maketitle

\section*{Introduction}
Composite structures formed by coupling a photonic micro/nano-resonator with an optical material supporting a strong exciton resonance provide unique capabilities to investigate and engineer light-matter interactions in solid-state systems \cite{kavokin2017microcavities,khitrova2006vacuum,baranov2018novel,gibbs2011exciton,vasa2018strong}.
The character of the interaction is largely determined by the ratio of the energy exchange rate between light and matter, i.e., the Rabi frequency, and the average dissipation rate.
In the weak coupling regime, the Rabi frequency is smaller than the average dissipation rate and the eigenstates of the structure can still be described in terms of the exciton and photon modes. Modifications of the local photonic density of states by the photonic structure can however alter the radiative life-time of the emitters
\cite{kinkhabwala2009large,akselrod2014probing,Hoang2015,Belacel2013}.
A qualitatively different situation arises in the strong coupling regime, where the Rabi frequency exceeds the dissipation rate and the energy is coherently exchanged between the light field and the emitters.
This interaction leads to the formation of new polaritonic eigenmodes with light and matter character \cite{kavokin2017microcavities}.
In the spectral domain, strong coupling manifests itself by an avoided crossing of the branches of the polaritonic eigenmodes
\cite{baranov2018novel,gibbs2011exciton,vasa2018strong}.

Achieving strong light-matter interactions requires a meticulous selection of both the photonic structure and the material system.
With regard to the photonic properties, the design of the structure must offer a suitable trade-off between the achievable vacuum field strength and the losses caused by absorption and radiation.
Strong coupling in solid state systems has been demonstrated utilizing high-Q dielectric microcavities \cite{reithmaier2004strong, PhysRevLett.95.067401}, photonic crystal cavities \cite{yoshie2004vacuum,hennessy2007quantum}, metallic microcavities \cite{Wang2016,Bisht2019}, surface plasmon polaritons \cite{torma2014strong,Gomez2010}, and various plasmonic nanocavities \cite{zengin2015realizing,liu2017strong,kleemann2017strong,leng2018strong,Wang2019}.

In terms of the material properties, the key requirements are an exciton state with large oscillator strength and low non-radiative damping rate. Moreover, the material should be robust and easy to combine with a wide range of photonic structures.  
Monolayers of transition metal dichalcogenides (TMDCs) such as tungsten disulfide (WS$_2$) meet these requirements \cite{schneider2018two}.
TMDC monolayers are atomically thin semiconductors with a direct band gap.
Their two-dimensional character in conjunction with the reduced screening by the dielectric environment results in exciton states with large oscillator strength and a binding energy of several hundred meV \cite{wang2018colloquium, AlexeyYang2018, He2014}.
The latter aspect renders TMDC monolayers an attractive material class for light-matter investigations at room temperature. 
Strong coupling of TMDC monolayer excitons and dielectric resonator \cite{liu2015strong,sidler2017fermi,Liu2014,Dufferwiel2015} and grating structures \cite{zhang2018photonic} as well as different plasmonic nanocavities \cite{Wang2016,wen2017room,geisler2019single,sang2021tuning} has been achieved.

\begin{figure*}[ht]
	\centering
	\includegraphics[scale=1]{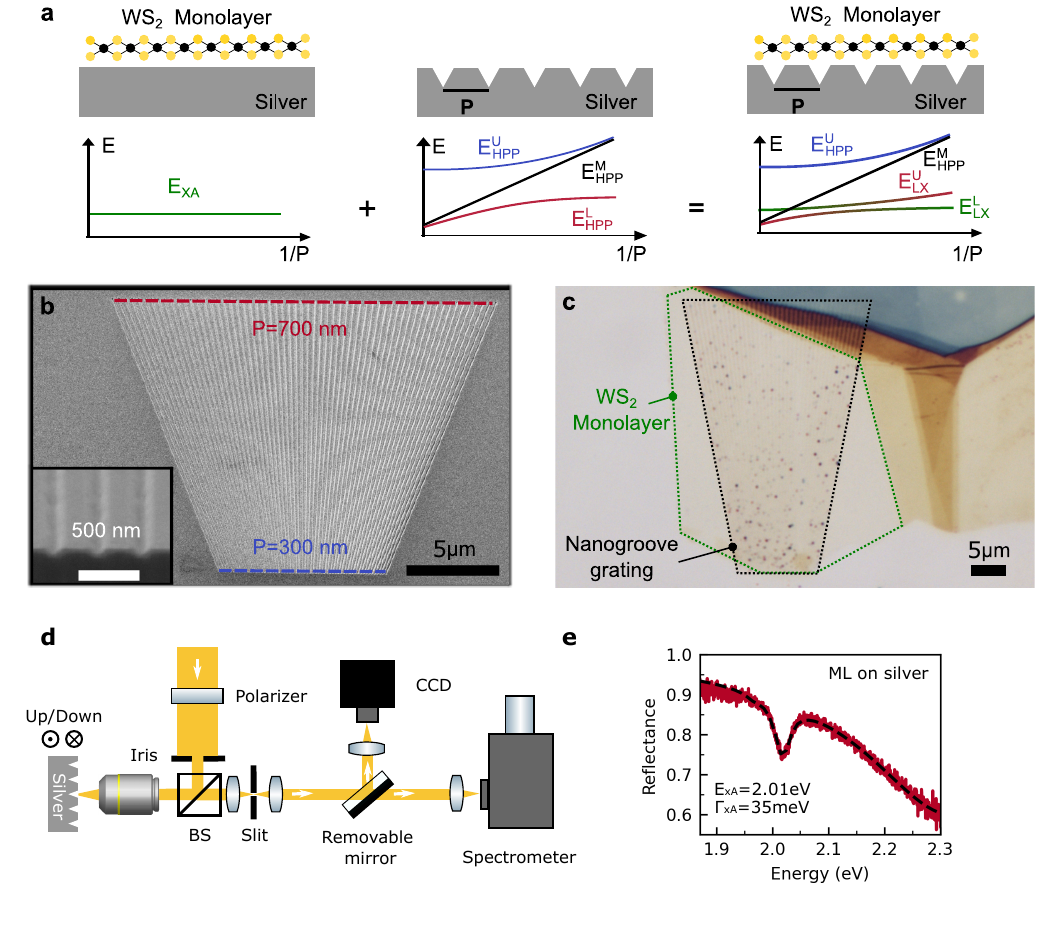}
	\caption{(a) Scheme of a composite structure consisting of a WS$_2$ monolayer deposited on top of a nanogroove grating milled into a single-crystalline silver flake  (upper panel). 
		The lower branch of the HPP of the nanogroove grating strongly couples with the A-exciton resonance of the monolayer resulting in an avoided crossing of the two modes (lower panel). 
		(b) Scanning electron micrograph of a fan-shaped nanogroove grating structure taken at a vertical tilt angle of $36^\circ$. The inset shows a crossection of the grating structure. (c) Optical micrograph of a WS$_{2}$ monolayer deposited on a nanogroove grating structure. The outlines of the monolayer and the grating are marked by the green and black dashed curve, respectively. (d) Scheme of the white light reflectance spectroscopy setup.
  (e) Reflection spectrum of a WS$_{2}$ monolayer deposited on a silver flake.}
	\label{PLot1}
\end{figure*}

Here, we report on a room-temperature study on the interaction of TMDC monolayer excitons with a hybrid plasmon polariton (HPP) mode.
For this purpose, we deposit WS$_2$ monolayers on nanogroove grating structures milled into  single-crystalline silver flakes (see Fig.\, \ref{PLot1}).
The nanogroove gratings host HPP modes that result from the strong coupling of localized surface plasmon resonances (LSPRs) in the grooves and propagating surface plasmon polaritons (SPPs) on the silver interface\cite{Wang2020,Yang2018}.
By engineering the geometry parameters of the nanogroove grating, we can control the HPP dispersion relation. In particular, we can vary the character of the HPP mode at the A-exciton energy from SPP-like to LSPR-like by increasing the nanogroove depth. 
Using reflection spectroscopy we demonstrate strong coupling between the A-exciton mode and the lower branch of the HPP for a SPP-like configuration.
Our results show the potential to tailor light-matter interactions in composite structures with a polaritonic sub-system.

\section*{Results and Discussion}

The composite structures were prepared according to the following procedure (see Methods for details).
In the first step, single-crystalline silver flakes with a typical lateral size of hundreds of microns and a thickness of several microns were grown on a silicon substrate using an ammonium hydroxide-controlled polyol reduction process~\cite{Wang2015}.
Next, focused ion beam milling was used to define fan-shaped nanogroove gratings in selected silver flakes (see Fig.\,\ref{PLot1} (b)). Within a grating structure, the period $P$ varies continuously from $300\,\mathrm{nm}$ (bottom) to $700\,\mathrm{nm}$ (top).
The depth $D$ of the nanogrooves was controlled by the milling time.
Finally, atomically thin WS$_2$ monolayers were prepared by a mechanical exfoliation method~\cite{Castellanos_Gomez_2014} and transferred onto the different nanogroove gratings. An optical micrograph of a completed sample is shown in 
Fig.\,\ref{PLot1} (c).

The samples were optically characterized with a home-built white light reflectance spectroscopy setup (see Fig.\,\ref{PLot1} (d)).
A halogen light bulb served as the light source. 
After reflection from a 50:50 beam splitter, the light was focused with a microscope  lens  (Mitutoyo M Plan Apo 50 NA=0.55) onto the sample. 
An iris diaphragm placed in front of the beam splitter was used to reduce the numerical aperture of the illumination to lower than 0.2.
The light reflected from the sample was collected with the same microscope lens and passed through the beam splitter.
An adjustable slit placed in an intermediate
image plane served as a spatial filter to select the portion of the reflected light from the fan-shaped grating corresponding to a specific grating period.
Reflection spectra for different grating periods were measured by shifting the fan-shaped grating structure along the nanogroove direction up or down.
The spectrum of the light was recorded with a cooled CCD camera (Princton Instruments PIXIS 256) attached to a spectrometer(Princton instruments ACTON SP 2300).

Before we address the composite structures, it is instructive to consider the optical properties of a WS$_2$ monolayer and the silver nanogroove gratings separately.
We start with the WS$_2$ monolayer.
 Fig.\,\ref{PLot1} (e) depicts the room temperature reflection spectrum of a WS$_2$ monolayer deposited on a planar silver flake. The spectrum is normalized with respect to the reflectance of the bare silver flake.
The spectrum features a pronounced dip at $E_\mathrm{xA}=2.01\,\mathrm{eV}$ that can be attributed to the A-exciton resonance of the monolayer~\cite{wang2018colloquium}. The full width at half maximum (FWHM) linewidth of the exciton resonance as obtained by a Lorentzian fit is $\Gamma_\mathrm{xA}=35\,\mathrm{meV}$, in line with recent reports on the exciton linewidth of a WS$_2$ monolayer coupled to a silver film \cite{sang2021tuning}.  
In comparison, the room temperature linewidth of WS$_2$ monolayers deposited on planar dielectric substrates is in the order of $25\,\mathrm{meV}$~\cite{selig2016excitonic}, indicating that the direct contact of the WS$_2$ monolayer with the single crystalline silver crystal only results in a moderate line broadening.

Next, we consider silver nanogroove gratings and the plasmonic modes supported by such structures.
Each nanogroove constitutes a nanoscale cavity that can host a LSPR.
Its resonance energy $E_{\rm{LSPR}}$ depends on the nanogroove geometry. For instance, increasing the 
nanogroove depth leads to a redshift of the resonance~\cite{sang2021tuning}.
If the nanogrooves are arranged in a periodic array, they form a grating coupler structure that allows to resonantly excite SPPs with an incident TM-polarized plane wave.
For normal incidence, the wave vectors of the excited SPPs follow from the phase matching condition 
$\bm{k}_{\rm{SPP}}  =\pm n \frac{2 \pi}{P} \hat{\bm{x}}$,
where  $P$ is the grating period, $\hat{\bm{x}}$ is the in-plane unit vector perpendicular to the nanogrooves, and $n$ is a positive integer. 

LSPRs and SPPs have different spectral characteristics.
SPPs at a flat silver-air interface exhibit an almost linear dispersion relation $E_ {\rm{SPP}}(\bm{k}_{\rm{SPP}})$ in the visible spectral range~\cite{novotny2012principles}.
In contrast, the LSPR of the nanogrooves is a dispersionless mode. Furthermore, the FWHM linewidth $\Gamma_{\mathrm{LSPR}}$ of the nanogroove LSPR  is expected to be significantly larger than the FWHM linewidth $\Gamma_{\mathrm{SPP}}$ of the SPP mode~\cite{Yang2018}. 

For suitable geometry parameters, the LSPR mode strongly interacts with the forward ($^+$) and backward ($^-$) propagating SPPs resulting in the formation of HPP modes~\cite{Wang2020}. 
The energy eigenvalues $E_{\mathrm{HPP}}$ of the  latter can be calculated using a coupled oscillator model $
\hat{\mathcal{H}} [ \alpha^+,	\beta,\alpha^-]^T=E_{\mathrm{HPP}}[ \alpha^+,	\beta,\alpha^-]^T$, 
where the Hamiltonian of the coupled system is given by
\begin{equation*}
	\hat{\mathcal{H}}=
	\begin{pmatrix}
		E_ {\rm{SPP}}^+-\imath\frac{ \Gamma_ {\rm{SPP}}^+}{2} & g& 0 \\ 
		g&E_{\rm{LSPR}}-\imath \frac{\Gamma_ {\rm{LSPR}}}{2}&g\\ 
		0 &g&E_{\rm{SPP}}^--\imath \frac{\Gamma_ {\rm{SPP}}^-}{2}
	\end{pmatrix}.
\end{equation*}
Here, $g$ is the coupling strength between the LSPR mode and the SPP modes. A direct coupling between the forward and backward moving SPPs is neglected in the model.
The Hopfield coefficients 
$\alpha^+$, 
$\alpha^-$, and $\beta$ specify the  SPP$^+$, SPP$^-$, and LSPR mode fractions, respectively, in the HPP modes, with $\vert\alpha^+\vert^2+\vert\alpha^- \vert^2+\vert\beta\vert^2=1$.
For normal incidence, the excited SPPs are energetically degenerate, i.e., the condition $E_ {\rm{SPP}}^+=E_ {\rm{SPP}}^-\equiv E_ {\rm{SPP}}$  holds.
Neglecting the imaginary parts, the energy eigenvalues of the upper, lower, and middle branch of the HPP are given by
\begin{eqnarray*}
E_{\rm{HPP}}^{\mathrm{U,L}}&=& \frac{E_{\rm{SPP}}+E_{\rm{LSPR}}}{2}\pm \frac{1}{2} \Delta E_\mathrm{UL},\\
	E_{\rm{HPP}}^\mathrm M&=& E_{\rm{SPP}},
\end{eqnarray*}
respectively. $\Delta E_\mathrm{UL}(\delta)=\sqrt{\delta^{2}+8 g^{2}}   $ is the energy splitting between the upper and lower branch, while $\delta=E_{\rm{SPP}}-E_{\rm{LSPR}}$ is the detuning between the SPP and LSPR modes.
For zero detuning,
the energy splitting is given by $\Delta E_\mathrm{UL}(0)=2 \sqrt{2} g$.
The FWHM linewidths of the polariton branches can be calculated with the corresponding Hopfield coefficients to $\Gamma_{\rm{HPP}}=\vert\alpha^+\vert^2 \Gamma_ {\rm{SPP}}+\vert\alpha^- \vert^2\Gamma_ {\rm{SPP}}+\vert\beta\vert^2\Gamma_ {\rm{LSPR}}$.

\begin{figure}[ht]
	\includegraphics[scale=1]{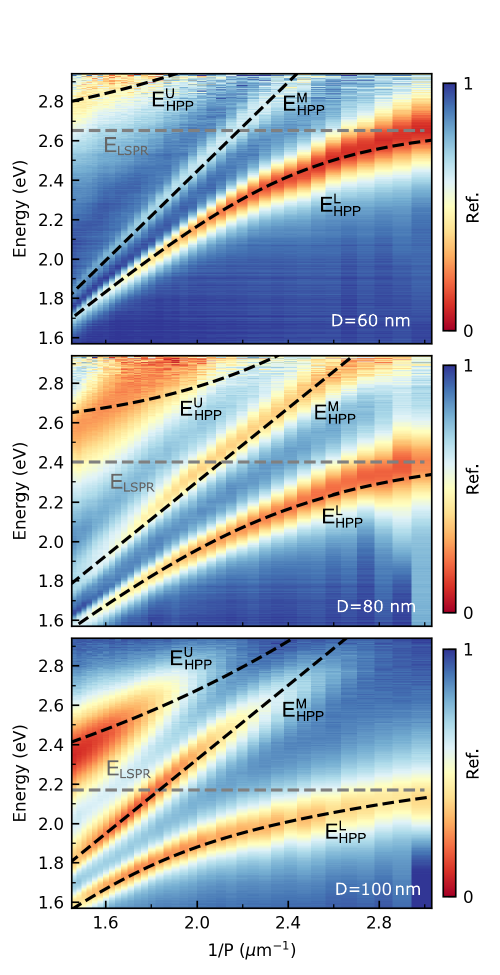}
	\caption{
		Normal incidence reflection spectra of  nanogroove grating structures with 60 nm (upper panel), 80 nm (middle panel), and 100 nm (lower panel) depth, respectively, recorded for TM polarized light. The dashed black curves are in each case fits to the data based on the coupled oscillator model.
		The horizontal gray dashed lines are the extracted resonance energies of the LSPR mode of the nanogrooves.}
	\label{Figure2}
\end{figure}

Figure \ref{Figure2} displays measured reflectance spectra versus the inverse grating period of the silver nanogroove gratings with depths of 60 nm, 80 nm, and 100 nm, respectively, recorded for TM-polarization (electric field vector perpendicular to the nanogrooves).
The sets of spectra feature in each case three bands of low reflectivity that we associate with the three HPP branches.
The dashed black curves are fits based on the above coupled oscillator model. For each set, the spectral position of the lower HPP branch for the smallest grating period was used to determine the LSPR resonance energy
$E_{\rm{LSPR}}$, while a linear fit of the middle HPP branch was used to determine $E_{\rm{SPP}}(1/P)$.
In the next step, the dispersion of the lower HPP branch was used to fit the coupling strength $g$.
While the model nicely reproduces the dispersion of the lower and middle HPP branch, the agreement is less good for the upper HPP branch.
This discrepancy can be attributed to some factors that are not included in the coupled oscillator model, e.g., the coupling of the LSPR mode to higher-order SPP modes and the effect of the first diffraction order of the nanogroove grating.  
As expected, the LSPR resonance energy $E_{\rm{LSPR}}$ shifts to lower energies as the nanogroove depth increases  (60 nm nanogrooves: $ 2.65\,\mathrm{eV}$; 80 nm nanogrooves: $2.4\,\mathrm{eV}$; 100 nm nanogrooves: $2.17\,\mathrm{eV}$).
In contrast, the zero detuning energy splitting $\Delta E_\mathrm{UL}(0)$ between the upper and lower polariton branch is comparable for the three nanogroove geometries (60 nm nanogrooves: $769\,\mathrm{meV}$; 80 nm nanogooves: $797\,\mathrm{meV}$; 100 nm nanogrooves: $730\,\mathrm{meV}$).

\begin{figure}[tt]
	\includegraphics[scale=1]{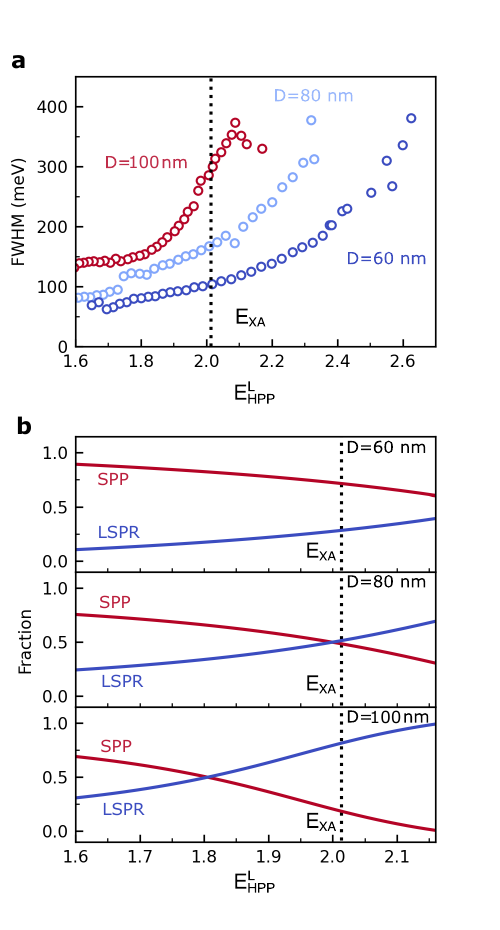}
	\caption{
		(a) Fitted linewidths of the lower HPP branch of the three nanogroove grating structures (dark blue: $D=60\,\mathrm{nm}$; light blue: $D=80\,\mathrm{nm}$; red: $D=100\,\mathrm{nm}$). (b) 
SPP- (red) and LSPR-mode (blue) fractions of the lower HPP branch of the three nanogroove grating structures as derived from coupled oscillator fits. The dashed black line is the extracted A-exciton energy of the WS$_2$ monolayer on a silver flake.}
	\label{Figure3}
\end{figure}

Figure \ref{Figure3} (a) shows the fitted linewidths $\Gamma_\mathrm{HPP}^\mathrm L$ of the lower HPP branch of the three nanogroove grating structures.  
For all three samples, we observe an increase of $\Gamma_\mathrm{HPP}^\mathrm L$ with $E_{\rm{HPP}}^\mathrm L$. 
This trend is in each case a consequence of the varying LSPR ($|\beta|^{2}$) and SPP ($|\alpha^+|^{2}+|\alpha^-|^{2}$) fractions of the respective lower HPP branch (see
Figure \ref{Figure3} (b)). 
For $E_{\rm{HPP}}^\mathrm L\ll E_{\mathrm{LSPR}}$, the lower HPP branch has a predominant SPP character and, hence, $\Gamma_\mathrm{HPP}^\mathrm L\approx\Gamma_ {\rm{SPP}}$.
As $E_{\rm{HPP}}^\mathrm L $ approaches $E_{\mathrm{LSPR}}$, the LSPR fraction and, hence, also $\Gamma_\mathrm{HPP}^\mathrm L$ increases.
For a fixed energy, this results in an increasing linewidth with the nanogroove depth.
For $D=60\,\mathrm{nm}$ and $E_{\rm{HPP}}^\mathrm L=E_{\rm{xA}}=2.01\,\mathrm{eV}$, the lower HPP branch has a predominant SPP character ($|\alpha^+|^{2}+|\alpha^-|^{2}=0.73$, $|\beta|^{2}=0.27$) resulting in a linewidth of $104\,\mathrm{meV}$.
In the case of the 80 nm deep nanogrooves, the SPP and LSPR fractions are comparable at the A-exciton energy ($|\alpha^+|^{2}+|\alpha^-|^{2}=0.49$, $|\beta|^{2}=0.51$) leading to an increase in linewidth to $168\,\mathrm{meV}$.
Finally, in the case of the 100 nm deep nanogrooves, the lower HPP branch is dominated by the LSPR contribution at the A-excition energy ($|\alpha^+|^{2}+|\alpha^-|^{2}=0.22$, $|\beta|^{2}=0.78$) and the line width is approximately $300\,\mathrm{meV}$.

\begin{figure*}[tttt]
	\includegraphics[scale=1]{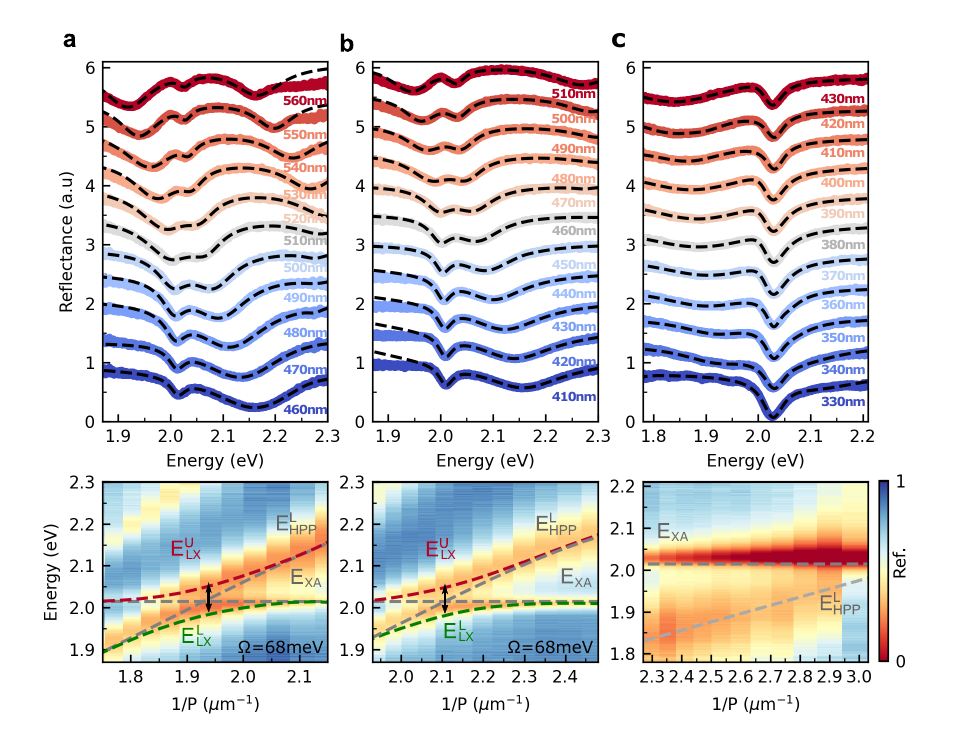}
	\caption{ (a)-(c) Normal incidence reflection spectra of WS$_{2}$ monolayer coupled to nanogroove grating structures with 60 nm, 80 nm, and 100 nm depth, respectively, recorded for TM polarized light. The lower HPP branch is near-resonance with the A-exciton mode. The black dashed lines in the upper panels fit the spectra with the multi-Lorentz model. The red and green dashed lines in the lower panels depict the upper and lower polariton branches based on the coupled oscillator model. The tilted and horizontal grey dashed lines are extracted resonance energies of the lower HPP branch and the A-exciton mode. }
 
	\label{Figure4}
\end{figure*}

Having characterized the individual components, we next discuss the properties of the composite structure. Fig.\,\ref{Figure4} (a) (top) depicts reflectance spectra  of a WS$_2$ monolayer deposited on the silver grating structure with 60 nm deep nanogrooves recorded with TM-polarized light.
The chosen grating periods range between $460\,\mathrm{nm}$ and $560\,\mathrm{nm}$ and the spectra are vertically offset for clarity.
For $P=460\,\mathrm{nm}$, the spectrum features two minima, that we identify as the
the A-exciton mode and the lower HPP branch.
The latter shows a red shift of $89\,\mathrm{meV}$ compared to the bare nanograting case due to the increase of the local refractive index at the silver interface caused by the WS$_2$ monolayer \cite{Shi2018}. 
As $P$ increases, we observe
a clear avoided crossing behavior of the two modes, indicating an interaction between the A-exciton and the lower HPP branch.
We note that a third reflection minimum observable for periods larger than $510\,\mathrm{nm}$ can be attributed to the middle HPP branch.

To model the interaction between the A-exciton mode and the lower HPP branch, we employ a coupled oscillator approach
 $
\hat{\mathcal{H}} [ \alpha_{\rm {HPP}},
\alpha_{\rm {XA}}]^T=E_{\mathrm{LX}}[ \alpha_{\rm {HPP}},
\alpha_{\rm {XA}}]^T$
with the Hamiltonian
\begin{equation*}
	\hat{\mathcal{H}} =\begin{pmatrix}
		E_{\rm{HPP}}^{\rm L}-\imath \frac{\Gamma_ {\rm{HPP}}^{\rm L}}{2} & \mbox{g}_{\rm{LX}}\\ 
		\mbox{g}_{\rm{LX}}&E_{\rm{XA}}-\imath \frac{\Gamma_ {\rm{XA}}}{2}
	\end{pmatrix}.
\end{equation*}
Here, $\mbox{g}_{\rm{LX}}$ is the coupling strength between the two modes.
The Hopfield coefficients 
$\alpha_{\rm {HPP}}$ and  
$\alpha_{\rm {XA}}$ specify the fractions of the lower HPP branch and the A-exciton mode, respectively. 
The effect of the middle and upper HPP branch are ignored in this model because of their large detuning with respect to the A-exciton mode for the chosen grating periods.
The complex eigenenergies of the upper and lower branch of the coupled system are
\begin{equation*}
	E_ {\rm{LX}}^{\rm{U,L}}=\frac{E_ {\rm{HPP}}^{\mathrm L}+E_ {\rm{XA}}}{2} -\imath\frac{\Gamma_ {\rm{HPP}}^{\mathrm L}+\Gamma_{\rm{XA}}}{4}\pm \frac{1}{2} \Delta E_\mathrm{UL},
\end{equation*}
where $\Delta E_\mathrm{UL}=2\sqrt{\mbox{g}_{\rm{LX}}^2+\frac{1}{4}\left[\delta_{\mathrm{LX}}-\frac{\imath}{2}(\Gamma_ {\rm{HPP}}^{\mathrm L}-\Gamma_{\rm{XA}})\right]^2}$  and 
$\delta_{\mathrm{LX}}={E_ {\rm{HPP}}^{\mathrm{L}}-E_ {\rm{XA}}}$ denotes the detuning between the lower HPP branch and the A-exciton mode. 
The energy difference between the upper and the lower branch for $\delta_\mathrm{LX}=0$, i.e., the Rabi splitting $\Omega$,  is given by $\Omega=2\sqrt{\mbox{g}_{\rm{LX}}^2-\frac{1}{16}\left[\Gamma_ {\rm{HPP}}^{\mathrm L}-\Gamma_{\rm{XA}}\right]^2}$.
The energies of the two interacting modes are determined for each period by fitting a multi-Lorentzian function to the corresponding reflection spectrum.
The dashed red and green lines superimposed on the color-coded reflectance spectra shown in Fig.\,\ref{Figure4} (a) (bottom), are the calculated dispersion of the upper and lower polariton branch, respectively, obtained by fitting the coupled oscillator model to the extracted mode energies.
From the measured data, we extract a Rabi splitting of $68\,\mathrm{meV}$.  
With $\Gamma_\mathrm{HPP}^\mathrm L=104\,\mathrm{meV}$ and $\Gamma_\mathrm{XA}=35\,\mathrm{meV}$, we 
obtain the coupling strength  $\mbox{g}_{\rm{LX}}=38\,\mathrm{meV}$. 
We compare this value with the critical coupling strength $\mbox{g}_\mathrm{c}=({\Gamma_{\rm{LSL}}+\Gamma_{\rm{XA}}})/{4}$ that  can be used to define  the boundary between the weak and strong coupling regime \cite{pelton2019strong, He2022, Cuadra2018,marko2022}.
From the measured linewidths, we infer a critical coupling strength $\mbox{g}_{c}$ of $35\,\mathrm{ meV}$. 
Since $\mbox{g}_{\rm{LX}}>\mbox{g}_{c}$, we claim that the composite structure with the 60 nm deep nanogrooves fulfills the strong coupling condition.

Next, we discuss the interaction of a WS$_2$ monolayer with the silver grating structure with 80 nm deep nanogrooves.
We anticipate that the larger nanogroove depth has two counteracting effects.
On the one hand, we expect that the reduction of the  LSPR energy leads to a stronger near-field and, thus, enhances the coupling strength $g_\mathrm{LX}$.
 On the other hand, the critical coupling strength $g_\mathrm{c}$ rises from $35\,\mathrm{ meV}$
 (60 nm nanogrooves) to $51\,\mathrm{ meV}$
 (80 nm nanogrooves) due to an increase of the linewidth of the lower HPP branch to  $\Gamma_\mathrm{HPP}^\mathrm L=168\,\mathrm{meV}$ at the A-exciton resonance energy (see discussion of  Fig.\,\ref{Figure3}).
 A priori, it is not clear which of these two effects dominates.
Fig.\,\ref{Figure4} (b) (top) shows reflectance spectra  of the composite structure with 80 nm deep nanogrooves for grating periods between $410\,\mathrm{nm}$ and $510\,\mathrm{nm}$ recorded with TM-polarized light. As in the previous case, we observe an avoided crossing of the A-exciton mode and the lower branch of the HPP.
The mode energies are determined by fitting a multi-Lorentzian function to the reflectance spectra and the calculated dispersion of the polariton branch is plotted versus the inverse grating period in Fig.\,\ref{Figure4} (b) (bottom).
The Rabi splitting $\Omega$ extracted from the experimental data is {$68\,\mathrm{meV}$}.
Together with $\Gamma_\mathrm{HPP}^\mathrm L=168\,\mathrm{meV}$ and $\Gamma_\mathrm{XA}=35\,\mathrm{meV}$, this results in a coupling strength $\mbox{g}_{\rm{LX}}$ of {$48\,\mathrm{meV}$}.
Since this value is smaller than the respective critical coupling strength $g_\mathrm{c}$, the composite structure with the 80 nm deep nanogrooves does not meet the strong coupling criterion, i.e., the increase of the damping dominates over the increase of the coupling strength.

One might now speculate that the situation could be reversed when the LSPR mode is brought into resonance with the A-exciton mode.
However, the reflectance spectra of the composite structure with 100 nm deep nanogrooves show that this does not happen (see Fig.\,\ref{Figure4} (c)).
The spectra feature two minima, a pronounced exciton resonance dip at $2.02\,\mathrm{eV}$ and a shallow dip caused by the lower HPP branch that shifts to higher energy with decreasing period.
As the lower HPP branch approaches the A-exciton, we do not observe an avoided crossing of the modes (also see Fig.\,\ref{Figure4} (c) (bottom)) and, hence, no indication for strong coupling.

\section*{Conclusion}
In summary, we have studied light-matter interactions at room temperature  in composite structures formed by coupling of WS$_2$ monolayers to different silver nanogroove gratings. 
The latter supports hybrid plasmon polaritons formed by strong-coupling of localized plasmon resonances in the nanogrooves with surface plasmon polaritons on the silver film.
By increasing the nanogroove depth from 60 nm to 100 nm, the character of the lower branch of the hybrid plasmon polariton at the energy of the WS$_2$ A-exciton mode changes from SPP-like to LSPR-like. 
This is accompanied by a significant increase in linewidth for this energy.
Using reflection spectroscopy, we demonstrate
strong coupling between the A-exciton mode and the lower branch of the HPP for the SPP-like configuration with 60 nm deep nanogrooves.
In contrast, the strong coupling criterion is not met for the LSPR-like configurations with the 80 nm and 100 nm deep nanogrooves.
This finding demonstrates the importance of taking into account both the achievable vacuum field strength and the damping during the design of the composite structure since a possible increase in the coupling strength can be easily overcompensated by larger damping.

We envision that those hybrid plasmon polaritons coupled to TMDC monolayers also hold great prospects for future investigations.
For instance, dark excitons in TMDC monolayers with zero in-plane transition dipole moment are challenging to detect with conventional far-field optical techniques but can be probed by near-field coupling to SPP modes \cite{Zhou2017}. 
The opportunity to tailor the field strength and distribution of the hybrid plasmon polaritons could be used in future experiments to tailor their interaction with dark excitons leading to efficient extraction of the dark exciton emission.
Moreover,  the potential to tailor the character of the HPP could also be interesting for the efficient  transfer of hot electrons into the TMDC monolayer \cite{Shan2019}.

\section*{Methods}
\subsection*{Synthesis of crystalline silver flakes} Monocrystalline silver crystals were synthesized based on an ammonium hydroxide-controlled polyol reduction process \cite{Wang2015}. Firstly,  we prepared a 0.17 M silver nitrate 15ml (Sigma Aldrich) ethylene glycol (EG) (Sigma Aldrich) solution. Then, ammonium hydroxide solution was added (28-30$\%$, Sigma Aldrich, 1.85 ml) to stabilize the reaction. In addition,  polyvinylpyrrolidone (Mw=55k, 0.5g), which acts as a capping agent, was added to slow down the dispersion as well as the rate of growth. Furthermore, chloroplatinic acid hydrate (H2PtCl6 · xH2O, $>$99.995, Sigma Aldrich, 0.54 mL of 0.02M in water) was added to form platin nanoparticles, which serve as nucleation centers for the silver atoms. Finally,  hydrogen peroxide (30$\%$, Chemsolute 1.3 mL) was added to start the reduction of the silver salt. The substrates are cleaned and added into the growth container before the mixture is added. Then the growth solution is left for several days. So the substrates are in the solution the whole time. After the growth, the substrates were cleaned with distilled water to remove excess chemicals. 

 \subsection*{Fabrication of nanogroove grating structures} Fan-shaped nanogroove gratings were fabricated by focus ion beam (FIB) lithography. The system is based on a Zeiss 1540XB Crossbeam microscope with a gallium (Ga2+) ion source and a Raith Elphy nanopatterning system. An ion beam with an acceleration voltage of 30 keV and a current of approximately 50 pA was used in combination with a 30 µm aperture. The relatively high current helps to reduce the patterning time to below 15 min.

 \subsection*{Preparation of WS$_2$ monolayers}  Atomically thin WS$_2$ monolayers were fabricated by a mechanical exfoliation method \cite{Castellanos_Gomez_2014} from a WS$_2$ crystal (2D Semiconductors). Monolayers were identified by micro differential reflectance spectroscopy\cite{Frisenda_2017, Niu2018}. The WS$_2$ monolayer was firstly deposited on a PDMS stamp and then transferred onto the nanograting structure. 
 
\section*{Funding}
This project was financially supported by the Deutsche
Forschungsgemeinschaft (DFG, German Research Foundation) under Germany’s Excellence Strategy – Cluster of
Excellence Matter and Light for Quantum Computing, ML4Q
(390534769).

\section*{Acknowledgements}
We thank A. Bergschneider and M. Wegerhoff for stimulating discussions.

\bibliography{bibstrongcoupling}

\end{document}